\documentclass[11pt]{amsart} 

\usepackage{amsmath,amssymb,amsthm,amscd,mathrsfs}
\usepackage[all]{xy}

\numberwithin{equation}{section}
\newcommand{\be}{\begin{equation}}
\newcommand{\ee}{\end{equation}}

\newcommand{\IP}{\mathbb{P}}%{{\relax{\rm I\kern-.18em P}}}

\newcommand\IZ{\mathbb {Z}}

\newcommand{\IC}{\mathbb{C}}

\newcommand{\IR}{\mathbb{R}}
\newcommand{\CU}{{\mathcal U}}
\newcommand{\ba}{\begin{array}}
\newcommand{\ea}{\end{array}}

\newcommand{\om}{\overline{M}}

\newcommand{\CV}{{\mathcal V}}

\newcommand{\bal}{\begin{aligned}}
\newcommand{\eal}{\end{aligned}}

\newcommand{\CZ}{{\mathcal Z}}

\newcommand{\half}{{1\over 2}}

\newcommand{\womega}{{\widetilde \omega}}
\newcommand{\wG}{{\widetilde G}}

\newcommand{\longto}{\longrightarrow}

\newcommand{\CO}{{\mathcal O}}

\newcommand{\oz}{{\overline z}}

\newcommand{\wJ}{{\widetilde J}}

\newcommand{\CT}{{\mathcal T}}

\newcommand{\oU}{{\overline U}}
\newcommand{\oV}{{\overline V}}
\newcommand{\wF}{{\widetilde F}}
\newcommand{\mcos}{{\mathrm{cos}}}
\newcommand{\msin}{{\mathrm{sin}}}
\title{Coisotropic Branes  in Toric Calabi-Yau 3-folds}
\author{N. Saulina}

\date{}

\begin{document}
\begin{titlepage}

\maketitle

%\vspace{1cm}

\centerline{\it Department of Physics and Astronomy}
\centerline{\it Rutgers University}
\centerline{\it Piscataway, NJ 08854-0849, USA}

\vspace{1.5cm}

 \begin{abstract} 
 \large
 We study disk-instantons ending on coisotropic branes preserved by real torus action in toric Calabi-Yau 3-folds. In particular, we find fermion zero modes on 
disk multi-covers ending on a coisotropic brane in local $\mathbb{P}^1$ geometry with
normal bundle $\mathcal{O}(-a)\oplus \mathcal{O}(a-2)$. 
It turns out that, independent of $a,$ disk multi-cover formula is the same
as for disks ending on a Lagrangian brane in resolved conifold.
We further construct an example of a coisotropic brane in Calabi-Yau 3-fold
 used in geometric engineering  of 4d $\mathcal{N}=2$
$SU(N)$ gauge theory, where this brane provides a surface
defect.
\end{abstract} 

\end{titlepage}

\tableofcontents

%\newpage

\section{Introduction}\label{intro}
Since the work \cite{KaOr}, it is known that boundary conditions in topological {\bf A}-model \cite{Witten} include not only Lagrangian branes but also
coisotropic branes. The latter are supported on higher than middle dimensional submanifolds of the symplectic target manifold and are characterized by nontrivial curvature 2-form of the connection of the complex line bundle on the world-volume.
Generalization to higher rank coisotropic branes was given in
\cite{Herbst}. The existence of coisotropic branes has important implications for
Homological Mirror Symmetry \cite{KaOr-II},\cite{KKOY}.

Computation of world-sheet instantons ending on  Lagrangian branes in {\bf A}-model is a well-developed subject. In \cite{AgVa},\cite{AgVaK}, the proof of mirror symmetry \cite{HV},\cite{HIV}
was used to provide  counting of holomorphic maps from genus zero Riemann
surface with boundary to holomorphic disks in toric 3-folds ending on Lagrangian branes.
In particular, the formula for disk multi-covers, predicted in \cite{OV}, was derived.
This formula was confirmed in \cite{LiuKatz,SongLi,GZ} by the virtual localization method. Later all genus
world-sheet instantons ending on Lagrangian branes  in toric Calabi-Yau 3-folds were summed
up \cite{AMV},\cite{AKMV}. This story was generalized for degenerate torus action
\cite{DFS}.

The subject of world-sheet instantons ending on coisotropic branes is much less explored.
So far explicit examples of coisotropic branes and topological strings ending on them were discussed in the special case of space-filling coisotropic brane \cite{KaW},\cite{GuW},\cite{Gu}.

In this note we study disks ending on coisotropic branes preserved by real torus action in toric Calabi-Yau 3-folds. In particular, we find fermion zero modes on 
disk multi-covers ending on a coisotropic brane in local $\mathbb{P}^1$ geometry with
normal bundle $\mathcal{O}(-a)\oplus \mathcal{O}(a-2)$. Surprisingly, the weights
of these fermion zero modes under the torus action are {\it exactly the same}
as for disk multi-covers ending on a toric Lagrangian brane
in resolved conifold computed in \cite{LiuKatz}.  These weights completely determine
the counting of holomorphic disk multi-covers, allowing 
 to write the
multiple-cover formula for disks ending on a coisotropic brane.

We further construct a coisotropic brane $Y$ in toric Calabi-Yau 3-fold
$\mathcal{X}_{SU(N)}$ which is used in geometric engineering \cite{KKV} of $\mathcal{N}=2,4d$
$SU(N)$ gauge theory in IIA string theory. By considering $D6$ brane supported on $R^{1,1}\times Y,$ one gets a surface defect in 4d theory with $(2,2)$ supersymmetric gauge theory in $R^{1,1}.$ We write
the multiple-cover formula for holomorphic disks in $\mathcal{X}_{SU(N)}$
 ending on $Y.$ This provides the leading contribution to the superpotential for the chiral field
 $y$ in 2d $(2,2)$ gauge theory.

This note is organized as follows. In Section 2 we review basic facts about 
disk instantons in topological $\bf{A}$-model.  In Section 3 we give a warm-up example of
a coisotropic brane in $\mathbb{C}^3.$ Section 4 provides construction of a coisotropic
brane in local $\mathbb{P}^1$ geometry with normal bundle $\mathcal{O}(-a)\oplus \mathcal{O}(a-2)$
and study of fermion zero modes on disk multi-covers
ending on this brane. In Section 5 we construct coisotropic brane in 
$\mathcal{X}_{SU(N)}$ and discuss its application as providing a surface defect
in 4d gauge theory.

\section{Disk instantons and {\bf A}-branes: Review}
Let us consider topological $\bf{A}$-model \cite{Witten} with K\"ahler target space $X$
and let $Y\subset X$ be a brane in this model.
Let $L$ be a complex line bundle on $Y$, 
$A$ a connection on $L$, and $F$ the curvature 2-form of $A$. 
Let $\beta \in H_2(X,Y)$, $\beta\neq 0$, and let 
$\om_{g,h}(X;Y,L,A;\beta)$ be a stable map compactification 
of the moduli space of holomorphic maps\footnote{Recall that path-integral of $\bf{A}$-
model localizes to holomorphic maps.} 
\[
f:(\Sigma, \partial\Sigma) \to (X,Y)
\]
from a genus $g\geq 0$ Riemann surface $\Sigma$ with $h\geq 1$ boundary components 
with boundary conditions specified by the triple $(Y,L,A)$, and $f_*[\Sigma]=\beta$. 
From a physical point of view, the data $(X,Y,L,A)$ determines a boundary topological ${\bf A}$-model \cite{KaOr}. The space 
$\om_{g,h}(X,Y,L,A,\beta)$ should be thought 
of as a  stable compactification 
of the moduli space of instantons in this {\bf A}-model coupled to topological 
gravity. 

The key in the path-integral approach to counting open world-sheet instantons
ending on a brane is to find fermion zero modes on a world-sheet with appropriate boundary
conditions specified by boundary matrix $R.$ 
Let $G$ denote the restriction of the K\"ahler metric on 
$X$ to $Y$. Note that the restriction $T_X|_Y$ of the 
complexified tangent bundle $T_X$ admits a 
direct sum decomposition (of $C^\infty$-bundles) 
\be\label{eq:tangdecomp}
T_X|_Y \simeq N_Y \oplus T_Y 
\ee
where $N_Y$ is the normal bundle to $Y$ in $X$, 
and $T_Y$ is the tangent bundle to $Y$. The boundary matrix $R: T_X|_Y\to T_X|_Y$ is the 
linear map defined by the following block matrix \cite{Polch}
\be\label{eq:bcmatrixA}
R =\left[\begin{array}{cc}
-1_{NY} & 0 \\
0 & (G-F)^{-1}(G+F)\\
\end{array}\right]
\ee 
with respect to the decomposition \eqref{eq:tangdecomp}.
%In untwisted $\sigma-$ model on $\Sigma$ wi the boundary conditions for fermions
%are 

In topological $\mathbf{A}$-model  on $\Sigma$ there are fermions 
 $$\chi \in \Gamma\Bigl(f^*\left(T_X^{1,0}\right)\Bigr)
 \quad \overline{\chi} \in \Gamma\Bigl(f^*\left(T_X^{0,1}\right)\Bigr)\quad
 \psi \in \Gamma\Bigl(\Omega^{0,1}_{\Sigma}\otimes f^*\left(T_X^{1,0}\right)\Bigr)\quad
 \overline{\psi} \in \Gamma\Bigl(\Omega^{1,0}_{\Sigma}\otimes f^*\left(T_X^{0,1}\right)\Bigr).$$
 They arise from fermions $\Psi_{\pm}\in \Gamma\Bigl(S_{\pm, \Sigma}\otimes TX\Bigr) $
 in untwisted $\sigma$-model as \cite{Witten}
 \be \label{relation} \begin{array}{c|c}
 {\text {\bf A}-model} & \sigma-{\text model}\cr
 \chi & \Psi_+^i \partial_{z_i}\cr
 \overline{\chi} & \Psi_-^{\bar i} \partial_{\bar z_i}\cr
 \psi & \Psi_-^i \partial_{z_i}\cr
 \overline{\psi} & \Psi_+^{\bar i} \partial_{\bar z_i}\cr
 \end{array}
 \ee
 where $z_i,\bar z_i$ are local complex coordinates on $X.$
 
 On $\partial\Sigma$ fermions satisfy boundary conditions
 \be\label{eq:boundcondA}
{\overline \chi}|_{\partial \Sigma} = \widetilde{R_+}(\chi|_{\partial \Sigma}) \qquad
{\overline \psi}|_{\partial \Sigma} = R_+(\psi|_{\partial \Sigma}).
\ee
 where $R_+$ and $\widetilde{R_+}$ are linear maps $R_+, \widetilde{R_+}: T_X^{1,0}|_Y\to T_X^{0,1}|_Y$ 
 determined by $R$.
Equations \eqref{eq:boundcondA} follow from boundary
 condition $\Psi_+=R(\Psi_-)$ on $\partial \Sigma$ in untwisted $\sigma$-model
 using relations \eqref{relation}.
 
Recall that in path-integral approach one uses the coupling $\int_{\Sigma} 
R_{m \bar i k \bar j} \,\chi^{m}\, \overline{\chi}^{\bar i}\,  \psi^{k} \wedge \overline{\psi}^{\bar j} $
in the action of $\mathbf{A}$-model to saturate fermion zero modes. (For closed $\Sigma$ this was done in \cite{AsMo}.) In this way one gets integral of the Euler class of certain vector bundle over the moduli space. For general $(X,Y)$ it is difficult to evaluate
this integral. However, in many cases of interest for physics one uses global symmetry
to localize this integral to a sum.
%virtual fundamental class $[\om_{g,h}(X,Y,L,A,\beta)]^{vir}.$

In this note we initiate the study of  open string instantons for coisotropic branes
preserved by the canonical (real) torus action in 
toric target spaces. Therefore it will be assumed in the following that $X$ is a toric Calabi-Yau threefold and that  5-dimensional coisotropic cycle 
$Y$ is preserved by the canonical $U(1)^3$-action on $X$. Moreover, $L$ is equipped 
with an 
equivariant structure so that the two-form $F$ is invariant under the torus action. 
When a rigorous construction of a moduli space equipped 
with a torus equivariant perfect tangent-obstruction theory is available, 
explicit computations rely on the virtual localization theorem \cite{GP}.
In certain cases of physical interest, these steps can be 
formally carried out even in the absence of a rigorous 
construction of a virtually smooth moduli space. Such an approach has been 
implemented for stable maps with Lagrangian boundary conditions 
in many examples \cite{LiuKatz, SongLi, GZ, DF, DFM, FLT}, the results being consistent with
mirror symmetry and large $N$ duality.

Let $z_i$, $1\leq i\leq 3$, be affine toric coordinates on a toric coordinate patch 
$U\subset X$, $U\simeq \IC^3$.
Suppose there is  a  holomorphic disk 
$D\subset U$ defined by the equations
\[
|z_1|\leq 1\qquad z_2=z_3=0
\]
so that $\partial D \subset Y$. 
As mentioned above, throughout this section it will be assumed that 
the cycle $Y$ and curvature 2-form $F$ are invariant under the canonical 
$U(1)^3$ action on $X$. In fact it suffices to work with a single one-parameter 
subgroup ${\bf T}\simeq U(1)\subset U(1)^3.$ 
 Then there is a natural induced action 
${\bf T}\times \om_{g,h}(X,Y,\beta)\to \om_{g,h}(X,Y,\beta)$ 
on the moduli space of open string instantons. This allows to evaluate
the integral of the Euler class over the moduli space as a sum of local contributions
from torus fixed points.

 Let $\Delta$ be the disk $|t|\leq 1$ in the complex 
$t$-plane and consider a degree $d\geq 1$ map $f:\Delta \to D$
of the form 
\[f(t)= (t^d,0,0).\]
Note that $[(\Delta,f)]$ 
is the unique torus invariant degree $d$ multi-cover of $D$ of type $(g,h)=(0,1)$ up to 
isomorphism. In particular it determines an isolated fixed point for 
the torus action on the moduli space of degree 
$d$ instantons of the topological {\bf A}-model defined by the data 
$(Y,L,A)$. Let $\CT_{D,R_+}$ ($\CT_{D,\widetilde{R_+}}$) be the sheaf of germs of holomorphic 
sections of the bundle $T^{1,0}_X|_D$ with boundary conditions \eqref{eq:boundcondA}.
Let $\CT_\Delta$ be the sheaf of germs of holomorphic sections of 
the tangent bundle $T^{1,0}_\Delta$ with natural real boundary conditions 
along $\partial \Delta$. 

The local contribution of the isolated fixed point $[(\Delta,f)]$ to the 
virtual localization formula is 
\be\label{eq:locEulerA} 
{1\over d}\,{e_{\bf T}\Bigl(H^1\left(\Delta, f^*\CT_{D,R_+}\right)\Bigr)\ e_{\bf T}\Bigl(H^0\left(\Delta, \CT_\Delta \right)\Bigr)\over 
e_{\bf T}\Bigl(H^0\left(\Delta,f^*\CT_{D,\widetilde{R_+}}\right)\Bigr)}.
\ee
where $e_{\bf T}$ denotes the equivariant Euler class. Recall that for a vector space $\mathbb{V}$
with {\bf T}-action, $e_{\bf T}(\mathbb{V})$ is the product of ${\bf T}$-weights of basic vectors.
One may think of
vector spaces $H^1\left(\Delta, f^*\CT_{D,R_+}\right)$ and $H^0\left(\Delta,f^*\CT_{D,\widetilde{R_+}}\right)$ as $\psi$ and $\chi$ zero modes respectively. 
The factor $e_{\bf T}\Bigl(H^0\left(\Delta, \CT_\Delta \right)\Bigr)$ takes into account
the $Aut$ group acting on $\Delta$

Once the local contribution of disk multi-covers are given, 
the local contribution of a generic torus fixed point 
in $\om_{g,h}(X,Y,\beta)$ is a standard exercise in 
virtual localization.

In \cite{LiuKatz, SongLi, GZ} the formula \eqref{eq:locEulerA} was used
for the case of Lagrangian branes.\footnote{with $R_+,\widetilde{R_+}$ which follow from $R=\begin{pmatrix} -1_{NY} & \cr
& 1_{TY}\cr \end{pmatrix}$} We use \eqref{eq:locEulerA} in Section 4 to compute
contribution of disk multi-covers ending on a coisotropic brane in
 the total space of 
a holomorphic rank two bundle of 
the form $\CO(-a)\oplus \CO(-b)$ over $\IP^1$, where $a+b=2$.

\section{A coisotropic {\bf A}-brane in $X=\IC^3$}\label{exampleIsect}
Our warm-up example is  a toric coisotropic {\bf A}-brane in $X=\IC^3$. As above, let 
$z_i$, $i=1,2,3$ be linear coordinates on $X$. 
The cycle $Y\subset X$ is given by $|z_1|=1$, therefore $Y\simeq S^1\times \IC^2$. 
The line bundle $L$ is trivial, and we set 
\be\label{eq:exampleIA} 
A = {1\over 2}\left(z_2d{ z_3} + \oz_2 d\oz_3\right),
\ee
which implies 
\be\label{eq:exampleIB}
F = {1\over 2}\left(dz_2\wedge dz_3 + d\oz_2\wedge d\oz_3\right).
\ee
Note that the restriction of the standard symplectic K\"ahler form 
\[
\omega = {i\over 2} \sum_{j=1}^3 dz_j \wedge d\oz_j
\]
to $Y$ is 
\[
\omega|_Y= {i\over 2} \sum_{j=2}^3 dz_j \wedge d\oz_j.
\]
Note that $Y$ is a coisotropic cycle in $X$ since it is real codimension 1. 
In order to check the remaining  conditions it is more convenient to use the 
real coordinates 
\[
z_1=r_1e^{i\theta_1}, \qquad z_2 = x_2 + i y_2, \qquad z_3=x_3+i y_3
\]
on the open subset $z_1\neq 0$, which covers $Y$. 
In terms of these coordinates
\be\label{eq:realformsA}
\bal
F & = dx_2\wedge dx_3 - dy_2 \wedge dy_3\\
\omega|_Y & = dx_2\wedge dy_2 + dx_3\wedge dy_3.\\
\eal 
\ee
The symplectic complement $T_Y^\perp$ of the tangent space 
$T_Y\subset T_X$ is spanned by $\partial_{\theta_1}$. 
Obviously, there is a natural direct sum decomposition 
\[
T_Y \simeq T_Y^\perp \oplus F_Y 
\]
where $F_Y\subset T_Y$ is spanned by $\partial_{x_i}, \partial_{y_i}$, $i=2,3$. 
Moreover, $\iota_{\partial_{\theta_1}}F=0$, therefore $F$ determines a section $\mathcal{F}$ of 
$\Lambda^2 F_Y^\ast$. Similary $\iota_{\partial_{\theta_1}}\omega=0$, hence $\omega$ 
determines a section $\sigma$ of 
$\Lambda^2 F_Y^\ast$. Let $J: F_Y\to F_Y$ be the linear map determined by 
\be\label{eq:transvcpxstrA} 
\mathcal{F}(h_1,h_2) = \sigma(h_1, J(h_2)) 
\ee
for any $h_1,h_2\in F_Y$.
Then a straightforward computation yields 
\be\label{eq:coiscondA}
J(\partial_{x_2})=-\partial_{y_3}\qquad J(\partial_{y_2})=-\partial_{x_3}\qquad 
J(\partial_{x_3})=\partial_{y_2}\qquad J(\partial_{y_3})=\partial_{x_2}.
\ee
This implies that $J^2=-1$, therefore $J$ is an almost complex structure on $F_Y$.

Finally, note that $Y$ is preserved by the {\bf T}-action on $X$ 
\be\label{eq:toractA}
e^{i\varphi}\times (z_j) \to (e^{i w_j \varphi} z_j), \qquad j=1,2,3, 
\ee
for any weights $w_j$. 
In order for 
$F$ to be invariant under the {\bf T}-action, the weights must satisfy $w_2=-w_3$. 

In order to compute the boundary matrix $R$ note that the K\"ahler metric $G$ is given by 
\be\label{eq:KahlerA}
G = dr_1\otimes dr_1 + r_1^2 d\theta_1\otimes d\theta_1 + \sum_{i=2}^3 
\left(dx_i\otimes dx_i+  dy_i\otimes dy_i\right)
\ee
in the above real coordinate chart. 
Therefore there is a natural $G$-orthogonal decomposition
\be\label{eq:ondecompA}
T_X|_Y \simeq N_Y \oplus T_Y^\perp \oplus F_Y 
\ee
where $N_Y$ is spanned by $\partial_{r_1}$ and $T_Y^\perp$ is the symplectic 
complement of $T_Y\subset T_X|_Y$, spanned by $\partial_{\theta_1}$. 
Then $R$ has the following block form 
with respect to the decomposition \eqref{eq:ondecompA}
\be\label{eq:RmatrixA}
R= \left[\begin{array}{ccc}
                                   -1 & 0 & 0 \\
                                    0 & 1 & 0\\ 
                                    0 & 0 & S \\
                                    \end{array}\right]
\ee
where $S:F_Y \to F_Y$ is the linear map determined by the condition 
\be\label{eq:RmatrixB} 
(G+F)(h_1,h_2) = (G-F)(h_1,S(h_2))
\ee
for any $h_1,h_2\in F_Y$. Using equations \eqref{eq:realformsA}, \eqref{eq:KahlerA}, 
a straightforward computation yields 
\be\label{eq:RmatrixC} 
S(\partial_{x_2}) = -\partial_{x_3}\qquad 
  S(\partial_{y_2}) = \partial_{y_3}\qquad 
S(\partial_{x_3}) = \partial_{x_2}\qquad 
S(\partial_{y_3}) = -\partial_{y_2}.
\ee                                  
In terms of holomorphic coordinate vector fields equations \eqref{eq:RmatrixA}, \eqref{eq:RmatrixC}
imply 
\be\label{eq:RmatrixD} 
R(\partial_{z_1}) = - e^{-2i\theta_1} \partial_{{\overline z_1}}\qquad 
R(\partial_{z_2}) = - \partial_{{\overline z_3}}\qquad 
R(\partial_{z_3}) = \partial_{{\overline z_2}}.
\ee
The remaining nontrivial $R$-matrix elements follow from \eqref{eq:RmatrixD} by conjugation.

\section{Coisotropic {\bf A}-branes in  local $\IP^1$ geometry}

In this section we present a more elaborate construction of a coisotropic 
{\bf A}-brane in the total space of 
a holomorphic rank two bundle of 
the form $\CO(-a)\oplus \CO(-b)$ over $\IP^1$, where $a+b=2$. 
We find fermion zero modes on holomorphic disk multi-covers ending on it and write a
formula counting these disks.
\subsection{Construction}\label{construction}
The total space $X$ of the rank two bundle $\CO(-a)\oplus \CO(-b)$ on $\IP^1$, 
$a,b\in \IZ_{\geq 0}$ 
admits a symplectic quotient construction $\IC^4//U(1)$. 
 In terms of complex linear coordinates $(X_1,X_2,U,V)$ the 
 symplectic form on $\IC^4$ is 
 \be\label{eq:sympformA} 
 \womega = {i\over 2} \left(dX_1 \wedge d{\overline X}_1 + dX_2 \wedge 
 d{\overline X}_2 + dU\wedge d\oU + dV \wedge d\oV\right)
 \ee
 and the symplectic $U(1)$ actions is given by 
 \be\label{eq:sympact} 
 e^{i\alpha} \times (X_1,X_2,U,V) \to \left(e^{i\alpha}X_1, e^{i\alpha}X_2, 
 e^{ia\alpha}U, e^{ib\alpha V}\right). 
 \ee
 The moment map 
of the symplectic $U(1)$-action is 
\be\label{eq:momentmap}
\mu(X_1,X_2,U,V) = |X_1|^2 + |X_2|^2 - a |U|^2 -b |V|^2.
\ee
For any $\zeta\in \IR_{>0}$, the level set
  $\CZ_\zeta=\mu^{-1}(\zeta)$ is a smooth manifold and the restriction of the 
  $U(1)$ action to $\CZ_\zeta$ is free. The quotient 
$X= \CZ_\zeta/U(1)$ is a smooth manifold equipped with a symplectic 
K\"ahler form $\omega$. Moreover there is a canonical 
principal $U(1)$-bundle structure $q: \CZ_\zeta\to X$ 
so that  
\be\label{eq:Kahlerdescent} 
{\widetilde \omega}|_{\CZ_\zeta} = q^* \omega.
\ee
By construction $X$ is isomorphic 
to the total space of the rank two bundle $\CO(-a)\oplus \CO(-b)$ on 
$\IP^1$, and the homogeneous toric coordinates $X_1, X_2$ are naturally identified 
with homogeneous coordinates on $\IP^1$. Let $\pi : X \to \IP^1$ denote the 
projection map. 

%Similarly, let ${\widetilde G}$ be the standard K\"ahler metric on $\IC^4$ and let $G$ 
%be the K\"ahler metric on $X$ with K\"ahler form $\omega$. 

Let $M\subset \CZ_\zeta$ 
be the codimension one cycle determined by the equation 
\be\label{eq:exampleIIA}
|X_2|^2=|X_1|^2 +c\quad c\ge 0
\ee
Obviously, $M$ is preserved by the symplectic $U(1)$-action. Since the 
later is free as observed above, the quotient $Y=M/U(1)$ is a codimension 
one cycle in $X$. Therefore $M$ is a coisotropic submanifold 
of $X$.  In fact equation \eqref{eq:exampleIIA} determines
a circle $S^1\subset \IP^1$, and $Y=\pi^{-1}(S^1)$. Therefore, again 
$Y\simeq S^1\times \IC^2$ as in Section 3. 
In the following we will determine a global vector field $\eta$ 
on $Y$ which generates the symplectic complement $T_Y^\perp 
\subset T_Y$ at any point. This vector field enters the definition
of coisotropic brane \cite{KaOr} as
$$ \mathcal{L}_{\eta}\,F=0,\qquad \mathcal{L}_{\eta}\left(\omega|_{Y}\right)=0.$$

First we define a suitable coordinate chart on $Z_\zeta$ covering $M$ and 
write down some explicit formulas for future reference. 
Given the moment map equation \eqref{eq:momentmap}, condition 
\eqref{eq:exampleIIA} implies that $X_1,X_2$ cannot vanish on $M$
as long as $a,b \in \mathbb{Z}_{\ge 0}$ and we fix parameter $\zeta$ such that $\zeta>c.$
Therefore $M$ is contained in the open subset $\CU\subset \CZ_\zeta$ 
defined by $X_1\neq 0$, $X_2\neq 0$.
The requirement of non-negativity of $a,b$ can be relaxed
when local $\mathbb{P}^1$ is part of more general geometry and there are
other reasons that $X_1,X_2$ cannot vanish on $M.$ We discuss examples 
with $a>0$ but $b<0$ in Section 5.

 The following real coordinate 
functions are well defined on $\CU$.
\be\label{eq:realcoordB}
X_j = r_j e^{i\theta_j},\quad j=1,2,\qquad U=U_1+iU_2\qquad V=V_1+iV_2.
\ee
In certain formulas it will be more convenient to use mixed coordinates of the form 
$(r_i,\theta_i, U,\oU, V, \oV)$, $i=1,2$. 
Then the K\"ahler form ${\widetilde \omega}$ and the moment map equation read
\be\label{eq:KahlerB}
{\widetilde \omega}|_\CU= r_1dr_1\wedge d\theta_1 + r_2dr_2\wedge d\theta_2 + 
{i\over 2} (dU \wedge d\oU + dV \wedge d\oV).
\ee
\be\label{eq:momentmapeqA} 
r_1^2 + r_2^2 - a|U|^2 -b|V|^2 =\zeta.
\ee
The restriction $T_{\CZ_\zeta}|_{\CU}$ of the tangent space to $\CZ_\zeta$ 
is isomorphic to the kernel of the differential 1-form 
\[  
\sum_{i=1}^2 (r_idr_i - aU_idU_i - bV_idV_i)
\]
on $T_{\IC^4}|_\CU$. An elementary computation shows that $T_{\CZ_\zeta}|_\CU$ is generated
by the vector fields 
\[
\left({\partial_{r_2}\over r_2}-{\partial_{r_1}\over r_1}\right)\qquad 
\partial_{\theta_i}\qquad 
\xi_i = \partial_{U_i} + {aU_i\over 2}\left({\partial_{r_1}\over r_1}+{\partial_{r_2}\over r_2}\right)\qquad \eta_i =  \partial_{V_i} + {bV_i\over 2}\left({\partial_{r_1}\over r_1}+{\partial_{r_2}\over r_2}\right)
\] 
with $i=1,2$. Note that these are well defined since $r_1\neq 0$, 
$r_2\neq 0$ in $\CU$. Since the defining equation of $M\subset \CU$ is $r^2_2=r_1^2+c$, 
it follows that there is a direct sum decomposition 
\be\label{eq:ondecompB} 
T_{\CZ_\zeta}|_M \simeq N_M \oplus T_M 
\ee
where the normal bundle $N_M$ is spanned by $(\partial_{r_2}/r_2-\partial_{r_1}/r_1)|_M$ 
and $T_M$ is spanned by $( \partial_{\theta_i}|_M, 
\xi_i|_M, \eta_i|_M)$, $i=1,2$.

Note also that 
equation \eqref{eq:momentmapeqA}
and the defining equation of $M$
yield 
\be\label{eq:restoM}
\bal 
{\widetilde \omega}|_M = 
 & {i\over 2} \left(dU\wedge d\oU +dV \wedge d\oV\right)
+{1\over 4} \left( aUd\oU + a\oU dU + bV d\oV +b \oV dV\right)\wedge d\theta_+
\eal
\ee
where $\theta_+ = \theta_1+\theta_2$. 
So that
\be \label{lie_omega} \mathcal{L}_{\partial_{\theta_-}}({\widetilde \omega}|_M) =0.\ee

The restriction of the symplectic $U(1)$-action to the neighborhood $\CU$ has the form 
\be\label{eq:gaugeact}
\bal 
e^{i\alpha} \times \left(r_1,r_2,\theta_1,\theta_2,U,\oU,V,\oV\right) \longto 
\left(r_1,r_2, \theta_1+\alpha, 
\theta_2+\alpha, e^{-ia\alpha}U,e^{ia\alpha}\oU, e^{-ib\alpha}V,e^{ib\alpha}\oV\right).
\eal
\ee
The canonical vector field  $\xi$ determined by the $U(1)$-action on $\CU$ is 
\be\label{eq:vectfield}
\xi = \partial_{\theta_1}+\partial_{\theta_2} + ia\left(\oU\partial_{\oU}-U\partial_U\right) + 
ib\left(\oV\partial_{\oV} -V\partial_V\right).
\ee
Note that 
\be\label{eq:contrA} 
\iota_\xi(\womega|_\CU) =0.
\ee

For future reference note that there is a canonical 
exact sequence 
\be\label{eq:exseqA} 
0\longto T_q \longto T_{Z_\zeta} {\buildrel p\over \longto} q^*T_X \longto 0
\ee
where $T_q$ is the vertical tangent bundle of the projection map $q:Z_\zeta\to X$.
At each point $z\in \CZ_\zeta$ the map $p_z:T_{\CZ_\zeta,z}\to 
(q^*T_X)_z$ is the differential of $q$ at $z$.  
By construction, the restriction $T_q|_{\CU}$ is spanned by the vector field $\xi$. 
Since $M$ is preserved by the $U(1)$-action \eqref{eq:gaugeact}, it follows that 
$T_q|_M\subset T_M$. 
Moreover, as $p$ maps $T_M \subset 
T_{\CZ_\zeta}|_M$ to $q^*T_Y$, there is an exact sequence 
\be\label{eq:exseqB} 
0\longto T_q|_M \longto T_M {\buildrel p\over \longto}
q^*T_Y\longto 0.
\ee
Equation \eqref{eq:Kahlerdescent} implies that
\be\label{eq:projformula} 
{\widetilde \omega}|_{\CZ_\zeta}( h_1,h_2) =(q^* \omega)(p(h_1),p(h_2))
\ee
for any two vectors $h_1,h_2\in T_{\CZ_\zeta}$. This implies in turn 
\be\label{eq:sympcomplA}
(q^*T_Y)^\perp = p(T_M^\perp)
\ee
where $(q^*T_Y)^\perp\subset q^*T_X|_M$ is the symplectic complement 
with respect to $q^*\omega$ and $T_M^\perp\subset T_{\CZ_\zeta}|_M$ 
is the complement with respect to ${\widetilde \omega}|_{\CZ_\zeta}$.

Since $M\subset \CU$ is defined by $r_2^2=r_1^2+c$, equations \eqref{eq:KahlerB}, \eqref{eq:contrA}
imply that $T_M^\perp$ is spanned by $\xi|_M$ and 
$(\partial_{\theta_2}-\partial_{\theta_1})|_M$. 
Therefore $(q^*T_Y)^\perp$ is spanned by $p(\partial_{\theta_2}-\partial_{\theta_1})|_M$.
The direct sum decomposition \eqref{eq:ondecompB} implies that 
$(\partial_{\theta_2}-\partial_{\theta_1})|_M$ belongs to $T_M$, hence 
$p(\partial_{\theta_2}-\partial_{\theta_1})|_M$ belongs to $q^*(T_Y)$. 
Moreover  $(\partial_{\theta_2}-\partial_{\theta_1})|_M$ 
is invariant under the $U(1)$-action \eqref{eq:gaugeact}, that is the Lie derivative 
with respect to $\xi|_M$ vanishes, 
\[
L_{\xi|_M} (\partial_{\theta_2}-\partial_{\theta_1})|_M=0.
\]
Therefore $p(\partial_{\theta_2}-\partial_{\theta_1})|_M$ descends to a nonvanishing vector 
field $\eta$ on $Y$ which generates $T_Y^\perp \subset T_Y$ at any point. 

Next we have to construct the line bundle $L$ with $U(1)$-connection $A$ on $Y$. 
We will take $L$ to be the trivial complex line bundle on $Y$. 
In order to construct the $U(1)$-connection on $L$, one can construct 
in principle a $U(1)$ connection on the trivial line bundle on $M$ which
descends to a connection on $L$ satisfying 
the coisotropic {\bf A}-brane conditions. Since $L$ is trivial it suffices to construct a 
closed real 2-form $\wF$ on $M$ which descends to a closed real 2-form 
$F$ on $Y$ so that $F$ satisfies the required conditions. 
Since there are no 
closed 2-cycles on $Y$, one can then find a globally defined connection 1-form 
$A$ on $Y$ so that $F=dA$. 

Invariance under $U(1)$ generated by $\xi\vert_M$ suggests the following ansatz
for $\wF$ 
\be\label{eq:exampleIIB}
{\wF} =  \half e^{i{(\theta_1+\theta_2)}}\Biggl( E_{(-a)}\wedge E_{(-b)}
+i \Bigl(\lambda_1 U E_{(-b)}+\lambda_2 V E_{(-a)}\Bigr)\wedge 
(d\theta_1-d\theta_2)\Biggr)+c.c.
\ee
where
$$E_{(-a)}=dU+iaUd\theta_1,\quad E_{(-b)}=dV+ibVd\theta_2$$
transform under the $U(1)$ in \eqref{eq:gaugeact} as
$$E_{(-a)}\mapsto e^{-i a \alpha}E_{(-a)}, \quad E_{(-b)}\mapsto e^{-i b \alpha}E_{(-b)}.$$
Now we impose\footnote{Recall the definition of Lie derivative $\mathcal{L}_v=\iota_v\,d +d \, \iota_v$} 
\be\label{eq:checkA}
d\wF =0 \qquad {\mathcal L}_{\partial_{\theta_-}}\wF =0 \qquad \theta_-= \theta_2-\theta_1
\ee
which fixes a linear combination of the two parameters in our ansatz:
\be \label{fix_lam}\lambda_1-\lambda_2=a-1\ee

Since $\wF$ is by construction preserved by the torus action \eqref{eq:gaugeact}, 
equations \eqref{eq:checkA} imply 
that there exists a closed real 2-form $F$ on $Y$ so that 
$\wF = q^*F$ and
\be\label{eq:contrC}
\mathcal{L}_{\eta}F =0
\ee
on $Y$. Recall that $\eta$ generates 
 the symplectic complement $T_Y^\perp \subset T_Y$, as shown below 
equation \eqref{eq:sympcomplA}.

Moreover, $\mathcal{L}_{\eta}(\omega|_Y) =0$ as a consequence of \eqref{lie_omega}
Therefore $F,\omega|_Y$ determine global  sections $\mathcal{F},\sigma$ of 
$\Lambda^2 F_Y^*$, where $F_Y= T_Y/T_Y^\perp$. Let $J: F_Y \to F_Y$ be the linear map determined by 
\[
\mathcal{F}(h_1,h_2) = \sigma(h_1,J(h_2))
\]
for any $h_1,h_2\in F_Y$. In order to complete the construction, one has to check 
that $J^2=-1$, hence $J$ defines an almost complex structure on $F_Y$. 

Let $F_M = T_M/T_M^\perp$, where $T_M^\perp\subset T_M$ is the  complement 
of $T_M$ in $T_{\CZ_\zeta}$ with respect to ${\widetilde \omega}|_{\CZ_\zeta}$. 
Since $T_M^\perp$ is spanned by $\xi|_M$ and $\partial_{\theta_-}|_M$, 
equations \eqref{lie_omega},\eqref{eq:checkA},  \eqref{eq:contrC} 
imply that $\wF,\womega|_M$ determine global sections 
$\widetilde{\mathcal{F}}, \widetilde{\sigma}$ of $\Lambda^2F_M^*$. Let ${\widetilde J}: F_M \to F_M$ be the linear 
map determined by 
\[
\widetilde {\mathcal{F}}(h_1,h_2) = {\widetilde \sigma}(h_1,{\widetilde J}(h_2))
\]
for any $h_1,h_2\in F_M$.

Now note that the exact sequence \eqref{eq:exseqB} and equation \eqref{eq:sympcomplA} 
imply that there exists a commutative diagram of the form 
\be\label{eq:commdiagA} 
\xymatrix{ 
 & 0\ar[d]& 0\ar[d] & & \\
0\ar[r] & T_q \ar[r] \ar[d]& T_q \ar[r]\ar[d]& 0 \ar[d]& \\
0 \ar[r] & T_M^\perp \ar[r]\ar[d]_-{p}& T_M \ar[r] \ar[d]_-{p}& F_M \ar[r]\ar[d] & 0\\
0 \ar[r] & q^*T_Y^\perp \ar[r] \ar[d]& q^*T_Y \ar[r]\ar[d] & q^*F_Y \ar[r] \ar[d]& 0\\
& 0  & 0  & 0 &  \\}
\ee
where all columns and rows are exact. In particular the pushforward map $p: T_M \to T_Y$ 
yields a natural isomorphism $F_M\simeq q^*F_Y$, which will also be denoted by $p$ 
in the following. Moreover, by construction 
\[
p\circ {\widetilde J} = (q^*J)\circ p.
\]
Therefore in order to prove that $J^2=-1$ it suffices to prove that ${\widetilde J}^2=-1$.
To this end, note that there is a direct sum decomposition 
\[
T_M \simeq T_M^\perp \oplus F_M 
\]
where $F_M$ is identified with the linear sub-bundle of $T_M$ spanned by 
the vector fields $(\xi_i|_M, \eta_i|_M)$, $i=1,2$. 
Recall that 
\[
\xi_i = \partial_{U_i} + {aU_i\over 2}\left({\partial_{r_1}\over r_1}+{\partial_{r_2}\over r_2}\right)\qquad \eta_i =  \partial_{V_i} + {bV_i\over 2}\left({\partial_{r_1}\over r_1}+{\partial_{r_2}\over r_2}\right)
\] 
for $i=1,2$. Equations \eqref{eq:exampleIIA}, \eqref{eq:restoM} imply that  
\[
\bal 
& \wF(\xi_i,\xi_j) = \wF(\partial_{U_i}, \partial_{U_j})\qquad 
\wF(\eta_i,\eta_j) = \wF(\partial_{V_i}, \partial_{V_j})\qquad 
\wF(\xi_i,\eta_j) = \wF(\partial_{U_i}, \partial_{V_j})\\
& {\womega}(\xi_i,\xi_j) = \womega(\partial_{U_i}, \partial_{U_j})\qquad 
\womega(\eta_i,\eta_j) = \womega(\partial_{V_i}, \partial_{V_j})\qquad 
\womega(\xi_i,\eta_j) = \womega(\partial_{U_i}, \partial_{V_j})\\
\eal
\]
for all $i,j=1,2$. Then a straightforward computation yields 
\be\label{eq:tildeJ} 
\bal 
& {\widetilde J}(\xi_1)= - \msin\theta_{ab}\, \eta_1-\mcos\theta_{ab}\, \eta_2\qquad 
\wJ(\xi_2) = -\mcos\theta_{ab}\, \eta_1 + \msin\theta_{ab}\, \eta_2 \qquad\\ 
& \wJ(\eta_1) = \msin\theta_{ab}\, \xi_1 +\mcos\theta_{ab}\, \xi_2\qquad 
\wJ(\eta_2) = \mcos\theta_{ab}\, \xi_1 -\msin\theta_{ab}\, \xi_2.\qquad\\
\eal
\ee
where $\theta_{ab}=a\theta_1+b\theta_2$. 
Equations \eqref{eq:tildeJ} easily imply $\wJ^2=-1$, concluding the construction. 

\subsection{Boundary matrix and disk multi-covers}\label{boundary}
As explained in Section 2, evaluation of disk amplitudes 
via localization requires an explicit expression for the boundary matrix $R$
in terms of local holomorphic coordinates on $X$. 
The standard affine toric coordinate patches on $X$ are 
\[
\bal
\CU_1: \quad {X_2}\neq 0\qquad z_1={{X_1}\over {X_2}} \qquad u_1 = X_2^aU \qquad v_1=X_2^bV \\
\CU_2:\quad {X_1}\neq 0 \qquad z_2={{X_2} \over {X_1}} \qquad u_2 = X_1^aU\qquad v_2=X_1^bV.\\
\eal 
\]
Note that $Y\subset \CU_1\cap \CU_2=q(\CU)$, where $\CU\subset \CZ_\zeta$ is the 
open subset $X_1\neq 0$, $X_2\neq 0$ introduced in the previous subsection. 

To simplify the computation of $R,$ we may set parameter $c=0$ in the defining equation
for $M$. The weights of fermion zero modes under the torus $\mathbf{T},$
and hence the contribution of the isolated fixed point to the virtual localization formula
\eqref{eq:locEulerA}, do not depend on $c.$

There are two torus invariant holomorphic disks in $X$ with boundary on $Y$ 
\[
\label{disks}
\bal
D_1:\qquad |z_1|\leq 1, \qquad u_1=v_1=0\\
D_2:\qquad |z_2|\leq 1, \qquad u_2=v_2=0\\
\eal
\]
It suffices to do the explicit computations only for $D_1$, since $D_2$ is entirely 
analogous. The boundary of $D_1$ is contained in both holomorphic coordinate 
charts $\CU_1,\CU_2$. Then a straightforward computation yields 
\be\label{eq:pushfwdA} 
\bal 
& q_*(X_1\partial_{X_2}|_M) = \partial_{z_2} \qquad
q_*(X_1^{-a} \partial_{U}|_M) = \partial_{u_2} \qquad 
q_*(X_1^{-b}\partial_V|_M) = \partial _{v_2}. \\
\eal 
\ee
Let us define real coordinate functions on $\CU_1\cap \CU_2$ by 
\be\label{eq:realcoordC}
z_2 = \rho_2e^{i\phi_2} \qquad u_2= x_2+iy_2 \qquad v_2=x_3+iy_3.
\ee
In terms of the real coordinates \eqref{eq:realcoordC}, 
\be\label{eq:pushfwdB} 
\bal 
& q_*(\partial_{r_1})|_{\partial D_1} = -{1\over r}\partial_{\rho_2}|_{\partial D_1} \qquad
q_*(\partial_{r_2})|_{\partial D_1} = {1\over r}\partial_{\rho_2}|_{\partial D_1}  \\
& q_*(\partial_{\theta_1})|_{\partial D_1} = -\partial_{\phi_2}|_{\partial D_1} \qquad 
q_*(\partial_{\theta_2})|_{\partial D_1}  = \partial_{\phi_2}|_{\partial D_1} \qquad \\
\eal 
\ee
\[
\bal
& q_*\left({1\over r^a} (\mcos(a\theta_1)\, \partial_{U_1} 
-\msin(a\theta_1)\, \partial_{U_2}) \right)\bigg|_{\partial D_1}  = \partial_{x_2}|_{\partial D_1} \\
 & q_*\left({1\over r^a} (\msin(a\theta_1)\, \partial_{U_1} 
+\mcos(a\theta_1)\, \partial_{U_2}) \right)\bigg|_{\partial D_1}  = \partial_{y_2}|_{\partial D_1}  \\
 & q_*\left({1\over r^b} (\mcos(b\theta_1)\, \partial_{V_1} 
-\msin(b\theta_1)\, \partial_{V_2}) \right)\bigg|_{\partial D_1}  = \partial_{x_3}|_{\partial D_1} \\
 & q_*\left({1\over r^b} (\msin(b\theta_1)\, \partial_{V_1} 
+\mcos(b\theta_1)\, \partial_{V_2}) \right)\bigg|_{\partial D_1}  = \partial_{y_3}|_{\partial D_1}  \\
 \eal 
\]
where $r=r_1=r_2$ on $\partial D_1$ satisfies the moment map equation $2r^2=\zeta$.
Note that the restriction $N_Y|_{\partial D_1}$ of the normal bundle to $Y$ 
in $X$ to the boundary of 
$D_1$ is generated by $\partial_{\rho_2}|_{\partial D_1}$. The restriction 
$T_Y|_{\partial D_1}$ is generated by $\partial_{\phi_2}|_{\partial D_1}, 
\partial_{x_i}|_{\partial D_1}, \partial_{y_i}|_{\partial D_1}$, $i=2,3$. 
The restriction of 
the symplectic complement $T_Y^\perp|_{\partial D_1}$
is generated by 
$\eta|_{\partial D_1} = \partial_{\phi_2}|_{\partial D_1}$. 
Therefore the restriction of the tangent space $T_X|_{\partial D_1}$ admits the direct sum decomposition 
\be\label{eq:bddecomp}
T_X|_{\partial D_1}\simeq N_Y|_{\partial D_1}\oplus T_Y^\perp|_{\partial D_1} 
\oplus F_Y|_{\partial D_1}
\ee
where $F_Y|_{\partial D_1}$ is generated by $\partial_{x_i}, \partial_{y_i}$, 
$i=2,3$. 

Rewriting equation \eqref{eq:exampleIIB} in real coordinates yields 
\be\label{eq:realwF} 
\wF =  \mcos(\theta_1+\theta_2)\, \Bigl((dU_1-aU_2 d\theta_1)\wedge (dV_1-bV_2 d\theta_2)-(dU_2+aU_1d\theta_1)\wedge (dV_2+bV_1d\theta_2)\Bigr) \ee
$$-
\msin(\theta_1+\theta_2) \, \Bigl((dU_1-aU_2 d\theta_1)\wedge (dV_2+bV_1d\theta_2) + (dU_2+aU_1d\theta_1) \wedge(dV_1-bV_2 d\theta_2) \Bigr) $$
$$-\lambda_1\, \mcos(\theta_1+\theta_2)\, \Bigl(U_1(dV_2+bV_1d\theta_2)+
U_2(dV_1-bV_2 d\theta_2)\Bigr)\wedge (d\theta_1-d\theta_2)$$
$$-\lambda_1\, \msin(\theta_1+\theta_2)\, \Bigl(U_1(dV_1-bV_2d\theta_2)-
U_2(dV_2+bV_1 d\theta_2)\Bigr) \wedge(d\theta_1-d\theta_2)$$
$$-\lambda_2 \,\mcos(\theta_1+\theta_2)\, \Bigl(V_1(dU_2+aU_1d\theta_1)+
V_2(dU_1-aU_2 d\theta_1)\Bigr)\wedge (d\theta_1-d\theta_2)$$
$$-\lambda_2 \,\msin(\theta_1+\theta_2)\, \Bigl(V_1(dU_1-aU_2d\theta_1)-
V_2(dU_2+aU_1 d\theta_1)\Bigr) \wedge(d\theta_1-d\theta_2)$$
where $\lambda_2 \in \mathbb{R}$ and $\lambda_1=a-1+\lambda_2.$
Then, using the identity 
\[
\wF(h_1,h_2) = F(q_*h_1,q_*h_2) 
\]
for any two tangent vectors $h_1,h_2\in T_M$, one obtains the following 
\be\label{eq:matrixelA}
\bal 
& F(\partial_{x_2},\partial_{y_2})|_{\partial D_1} = 0 \qquad 
F(\partial_{x_3},\partial_{y_3})|_{\partial D_1} = 0 \qquad \\
& F(\partial_{x_2},\partial_{x_3})|_{\partial D_1} = {1\over r^{2}} 
\mcos(\phi_2) \qquad 
F(\partial_{y_2},\partial_{y_3})|_{\partial D_1} = - {1\over r^{2}} 
\mcos(\phi_2) \qquad \\
& F(\partial_{x_2},\partial_{y_3})|_{\partial D_1} = -{1\over r^{2}}
\msin(\phi_2)\qquad 
F(\partial_{x_3},\partial_{y_2})|_{\partial D_1} = {1\over r^{2}}
\msin(\phi_2).\\
\eal 
\ee
\[
F(\partial_{\phi_2}, \partial_{x_i})|_{\partial D_1} =F(\partial_{\phi_2}, \partial_{y_i})|_{\partial D_1} =0
\]
for $i=2,3$. 

In order to compute the boundary matrix $R$, we also have to evaluate the 
quotient K\"ahler metric $G$ on the coordinate vector fields  $\partial_{\phi_2}, \partial_{x_i},\partial_{y_i}$, $i=2,3$, restricted to ${\partial D_1}$. 
The local expression of the K\"ahler metric ${\widetilde G}|_{\CU}$ is 
\be\label{eq:locmetricA} 
{\widetilde G}|_\CU = 
\sum_{i=1}^2 (dr_i\otimes dr_i + r_i^2 d\theta_i\otimes d\theta_i )|_{\CU}
+ \sum_{i=1}^2 (dU_i\otimes dU_i+dV_i\otimes dV_i)|_{\CU}.
\ee
The inverse image  $q^{-1}({\partial D_1})\subset \CU$ is determined by the 
equations 
\[
r_1=r_2 \qquad U=V=0.
\]
Therefore 
\[
\xi|_{q^{-1}({\partial D_1}) } = 
(\partial_{\theta_1}+\partial_{\theta_2})|_{q^{-1}({\partial D_1}) } 
 = 2\partial_{\theta_+}|_{q^{-1}({\partial D_1}) }
 \]
Then the orthogonal complement of $\xi|_{q^{-1}({\partial D_1}) }$ in 
$T_M|_{q^{-1}({\partial D_1}) }$ with respect to the metric $\wG|_M$ 
is generated by 
\[
\partial_{\theta_-}|_{q^{-1}({\partial D_1}) }, \qquad 
\partial_{U_i}|_{q^{-1}({\partial D_1}) }, \qquad 
\partial_{V_i}|_{q^{-1}({\partial D_1}) },\qquad i=1,2, 
\]
where $\theta_-=\theta_2-\theta_1$ and $\partial_{\theta_-} = {1\over 2}(\partial_{\theta_2}-\partial_{\theta_1})$. Moreover, note that the 
induced metric on the orthogonal complement of $\xi$ is invariant under the
$U(1)$ action. Therefore, using relations \eqref{eq:pushfwdB} one obtains 
\be\label{eq:matrixelB} 
\bal 
G(\partial_{x_2},\partial_{x_2})|_{\partial D_1} =  G(\partial_{y_2},\partial_{y_2})|_{\partial D_1}={1\over r^{2a}},\qquad
G(\partial_{x_3},\partial_{x_3})|_{\partial D_1} =  G(\partial_{y_3},\partial_{y_3})|_{\partial D_1}={1\over r^{2b}},
\eal 
\ee 
$$G(\partial_{\phi_2}, \partial_{\phi_2})|_{\partial D_1}  = r^2,$$

all other matrix elements of $G$ being trivial. 

Equations \eqref{eq:matrixelA}, \eqref{eq:matrixelB} 
imply that the linear map $R|_{\partial D_1} : T_X|_{\partial D_1} \to 
T_X|_{\partial D_1}$ has the following block form with respect to the 
direct sum \eqref{eq:bddecomp}
\be\label{eq:boundarymatrixA} 
R|_{\partial D_1} = \left[\begin{array}{ccc} 
-1 & 0 & 0 \\
0 & 1 & 0 \\
0 & 0 & S \\
\end{array}\right]
\ee 
where $S:F_Y|_{\partial D_1}\to F_Y|_{\partial D_1}$ 
is determined by the following condition
\[
(G+F)(h_1,h_2)|_{\partial D_1} = (G-F)(h_1, S(h_2))|_{\partial D_1} 
\]
for any $h_1, h_2\in F_Y|_{\partial D_1}$. Then a straightforward computation yields
\[
\bal 
S(\partial_{x_2}) &= r^{2-2a}\Bigl(-\mcos(\phi_2)\, \partial_{x_3} + \msin(\phi_2)\, 
\partial_{y_3}\Bigr)\\
S(\partial_{y_2}) &= r^{2-2a}\Bigl(\msin(\phi_2)\, \partial_{x_3} + \mcos(\phi_2)\, \partial_{y_3}\Bigr)\\
S(\partial_{x_3}) &=r^{2-2b} \Bigl(\mcos(\phi_2)\, \partial_{x_2} - \msin(\phi_2)\, \partial_{y_2}\Bigr)\\
S(\partial_{y_3}) &= r^{2-2b}\Bigl(-\msin(\phi_2)\, \partial_{x_2} - \mcos(\phi_2)\, \partial_{y_2}\Bigr)\\
\eal 
\]
where recall that $r^2={\zeta \over 2}$ on ${\partial D_1}.$
In terms of holomorphic coordinate vector fields we obtain 

\be\label{eq:boundarymatrixB} 
R({\partial z_2})|_{\partial D_1}  = -  e^{-2i\phi_2} \partial_{\overline z_2}|_{\partial D_1}
\quad 
R({\partial u_2})|_{\partial D_1} =  -e^{i\phi_2}\, r^{2-2a}\, \partial_{\overline v_2}|_{\partial D_1}\quad
R({\partial v_2})|_{\partial D_1} =  e^{i\phi_2} \, r^{2-2b}\, \partial_{\overline u_2}|_{\partial D_1}
\ee
and complex-conjugated equations.

Let $f:\Delta \to X$ be a degree $d\geq 1$ 
torus invariant holomorphic map with coisotropic boundary 
conditions along $Y$ which factors through the embedding $D_1\subset X$. 
%of the cohomology spaces $H^{i}(\Delta, \CT_\Delta)$, $H^i(\Delta, f^*\CT_{D,R_+})$, 
%$i=1,2$, defined in section (\ref{diskmulticover}).

 Let $\CV_1,\CV_2\subset \Delta$ be 
an  open cover of $\Delta$ 
so that $\CV_1=\Delta \setminus \partial_\Delta$, and $\CV_2=\Delta \setminus \{0\}$. 
Let $t_1,t_2$ be affine coordinates on $\CV_1,\CV_2$ 
so that the map $f$ is locally given by 
\be\label{eq:localmapA-ii} 
z_1= t_1^d\qquad z_2=t_2^d.
\ee
Note that $t_1$ is centered at the origin and $t_1t_2=1$ 
on the 
overlap $\CV_1\cap \CV_2$.

As reviewed in Section 2, $R$ determines boundary conditions for fermions 
\be\label{bry_psi}\Psi_+\vert=R(\Psi_-\vert)\ee
where by $\vert$ we mean $\vert_{\partial \Delta}$ and in the coordinate chart $\mathcal{U}_2$
$$\Psi_{\pm}=\Psi_{\pm}^{z_2}\partial_{z_2}+\Psi_{\pm}^{u_2}\partial_{u_2}+\Psi_{\pm}^{v_2}\partial_{v_2}+\Psi_{\pm}^{\bar z_2}\partial_{\bar z_2}+\Psi_{\pm}^{\bar u_2}\partial_{\bar u_2}+\Psi_{\pm}^{\bar v_2}\partial_{\bar v_2}.$$
In components, (\ref{bry_psi}) implies the following boundary conditions
$$\Psi_+^{z_2}\vert=-e^{2i\phi_2} \Psi_-^{\bar z_2}\vert,\quad
\Psi_+^{\bar z_2}\vert=-e^{-2i\phi_2} \Psi_-^{z_2}\vert,\qquad
\Psi_+^{u_2}\vert=e^{-i\phi_2} r^{2-2b}\,\Psi_-^{\bar v_2}\vert,$$
$$\Psi_+^{\bar u_2}\vert=e^{i\phi_2} r^{2-2b}\,\Psi_-^{v_2}\vert,\quad 
\Psi_+^{v_2}\vert=-e^{-i\phi_2} r^{2-2a}\,\Psi_-^{\bar u_2}\vert,\quad
\Psi_+^{\bar v_2}\vert=-e^{i\phi_2} r^{2-2a}\, \Psi_-^{u_2}\vert$$
where $r^2\vert={\zeta\over 2}$ and, as follows from \eqref{eq:localmapA-ii}, $e^{i\phi_2}\vert=t_2^d\vert.$ 
Now we use relation \eqref{relation} between $\Psi_{\pm}$ and fermions in $\bf{A}$-model
where in the present case  $i=z_2,u_2,v_2,$ and we find
boundary conditions \footnote{In untwisted $\sigma$-model reality condition on fermions
$\left(\Psi_{\alpha}^I\right)^*=\epsilon_{\alpha \beta} J^I_K\Psi_{\beta}^K$ uses
almost complex structure $J$ on the target space. 
In topological {\bf A}-model one can choose any $J.$ We take $J$ such that
 fermionic  reality conditions are $(\chi^{u_2})^*=\chi^{\bar u_2},\, (\chi^{v_2})^*=-\chi^{\bar v_2},\,(\chi^{z_2})^*=\chi^{\bar z_2},\quad (\psi^{u_2})^*=\psi^{\bar u_2},\, 
(\psi^{v_2})^*=-\psi^{\bar v_2},\,  (\psi^{z_2})^*=\psi^{\bar z_2}.$ }
 for $\chi$
\be \label{bry_0_form} \chi^{z_2}\vert=-e^{2i\phi_2}\chi^{\bar z_2}\vert,\quad
 \chi^{u_2}\vert=e^{-i\phi_2}r^{2-2b}\,\chi^{\bar v_2}\vert,\quad
  \chi^{v_2}\vert=-e^{-i\phi_2}r^{2b-2}\,\chi^{\bar u_2}\vert\ee
  and for $\psi$
  \be \label{bry_1_form} \psi^{z_2}\vert=-e^{2i\phi_2}\psi^{\bar z_2}\vert,\quad
 \psi^{u_2}\vert=-e^{-i\phi_2}r^{2-2b}\psi^{\bar v_2}\vert,\quad
  \psi^{v_2}\vert=e^{-i\phi_2}r^{2b-2}\,\psi^{\bar u_2}\vert\ee

In Appendix A.1 we found $\chi$ zero modes
  \be \label{chi_zero}\chi^{(zm)}=\Bigl(\sum_{m=0}^{2d}\alpha'_m t_2^m \Bigr) \partial_{z_2},\quad \overline{\alpha'_m}=-\alpha'_{2d-m}.\ee

 If  $\mathbf{T}$ acts on $X$ as 
\be \label{globalu1}X_1 \mapsto e^{i \varphi}X_1,\quad X_2 \mapsto X_2,\quad
U \mapsto e^{i n \varphi}U,\quad V \mapsto e^{-i(n+1)\varphi }V\ee
then we find the weights of $\chi^{(zm)}$:
\be \label{weights_chi} \{{1\over d},\ldots, {d-1\over d}, 1\} \quad \& \quad 0_R\ee
where  $0_R$ corresponds to the real mode with $m=d$ in (\ref{chi_zero}) and
weights in the bracket correspond to complex modes for $m=0,\ldots, d-1.$

In Appendix A.2 we found $\psi$ zero modes (which exist only if $d>1$)
\be \label{zero_psi}\psi^{(zm)}=\sum_{k=1}^{d-1}
\Bigl(b_k t_2^{k-ad}\partial_{u_2}+c_k t_2^{k-bd}\partial_{v_2}\Bigr)\quad b_k=-r^{2-2b}\,\overline{c_{d-k}}\quad d>1.\ee

The weights of $\psi^{(zm)}$ under \eqref{globalu1} are
\be \label{weights_psi} \{n+{1\over d},\ldots, n+{d-1\over d}\}.\ee
In  \eqref{weights_chi} and \eqref{weights_psi}  we recognize the weights of
 fermion zero modes for disk multi-covers
ending on a toric Lagrangian  brane in resolved conifold \cite{LiuKatz}. Then the localization formula \eqref{eq:locEulerA}  gives the 
contribution of the multi-covers of $D_1$ 
\be \label{multiple}\mathcal{W}=\sum_{d=1}^{\infty}{N_d(n)\over d^2}e^{dy} e^{-dt/2}\qquad N_1=1\quad N_d(n)={\prod_{j=1}^{d-1}(j+nd)\over (d-1)!} \in \mathbb{Z} \quad \text{for}\quad d>1.\ee
In  \eqref{multiple} $y=c+i\mathcal{A}$ 
where $c$ is a parameter in $|X_2|^2=|X_1|^2+c$ and $\mathcal{A}$ is a Wilson line
around $\partial D_1$ in $Y,$ and $t$ is complexified K\"ahler modulus of the local geometry.
 Note that \eqref{multiple} is independent of $a$ specifying the normal bundle $\mathcal{O}(-a)\oplus \mathcal{O}(a-2)$
but depends on $n$ which enters in the choice of $\bf{T}$-action \eqref{globalu1}.

So far we discussed multi-covers of $D_1$. One can do an analogous
computation for $D_2,$ the second torus invariant holomorphic disk in $X$ with boundary on $Y.$ One finds fermion zero modes
 \be \label{chi_zero_ii}\chi^{(zm)}=\Bigl(\sum_{m=0}^{2d}\alpha'_m t_1^m \Bigr) \partial_{z_1},\quad \overline{\alpha'_m}=-\alpha'_{2d-m}.\ee
\be \label{zero_psi_ii}\psi^{(zm)}=\sum_{k=1}^{d-1}
\Bigl(b_k t_1^{k-ad}\partial_{u_1}+c_k t_1^{k-bd}\partial_{v_1}\Bigr)\quad b_k=-r^{2-2b}\,\overline{c_{d-k}}.\ee
and the 
contribution of the multi-covers of $D_2$ is
\be \label{multiple_opp}\mathcal{W}=\sum_{d=1}^{\infty}{N_d(\tilde n)\over d^2}e^{-dy}e^{-dt/2}\ee
 where $\tilde n=n+1-b.$
  
 \section{Surface defect in geometrically engineered $SU(N)$ gauge theory}
 Here we give an example of coisotropic brane $Y$ in toric Calabi-Yau 3-fold $\mathcal{X}_{SU(N)}$
 used in IIA string theory for geometric engineering \cite{KKV}  of 4d $\mathcal{N}=2$ $SU(N)$ gauge theory. By considering $D6$ brane supported on $R^{1,1}\times Y,$ one gets a surface defect in 4d theory with $(2,2)$ supersymmetric gauge theory on $R^{1,1}.$
Similar to \cite{BCOV},\cite{OV},\cite{KKLM}, counting disk multi-covers in $\mathcal{X}_{SU(N)}$
 ending on $Y$ gives contribution to the superpotential for the chiral field
 $y$ supported on the surface defect. 
 
Let us first consider $SU(2)$ gauge theory arising from $\mathcal{X}_{SU(2)}$ defined
as a toric manifold
\be \label{action}\begin{array}{c|c|c|c|c|c}
&X_1 & X_2 & X_3  & X_4 &W \cr
\hline
\mathbb{C}_{(1)}^*&1 & 1 & 1& 0& -3 \cr
\hline
\mathbb{C}_{(2)}^*&0& 0& 1& 1& -2\cr
\end{array}
\ee
The construction starts from defining  $M$ as real codimension 3 submanifold
in $\mathbb{C}^5$
$$|X_1|^2+|X_2|^2+|X_3|^2-3|W|^2=\zeta_1,\quad
|X_3|^2+|X_4|^2-2|W|^2=\zeta_2,\quad |X_2|^2=|X_1|^2+c.$$
To ensure
that $X_1,X_2\ne 0$ on $M$ and hence $\theta_1$ and $\theta_2$ are well-defined, we must choose  $\zeta_1$ and $\zeta_2$ such that $\zeta_1-\zeta_2-c>0.$
$Y$ is defined as a quotient of $M$ by the two symplectic $U(1)$ actions
$$\xi_1: \{X_1,X_2,X_3,X_4,W\}\mapsto \{e^{i\alpha} X_1,e^{i\alpha}X_2,e^{i\alpha}X_3,X_4,e^{-3i\alpha}W\}$$
$$\xi_2: \{X_1,X_2,X_3,X_4,W\}\mapsto \{X_1,X_2,e^{i\beta}X_3,e^{i\beta} X_4,e^{-2i\beta}W\}$$

The following closed 2-form is invariant under both $\xi_1$ and $\xi_2$ actions
$$\tilde F=e^{i(\theta_{1}+\theta_2)} \Biggl((X_3dX_4-X_4dX_3)\wedge dW-2WdX_3\wedge dX_4-i\,X_4(X_3dW+2WdX_3)\wedge d\theta_1$$
$$+i\, W (X_3dX_4-X_4dX_3)\wedge d\theta_2-WX_3X_4 d\theta_1\wedge d\theta_2\Biggr)+c.c$$
and gives rise to the closed 2-form $F$ on $Y.$ Let us clarify that $\tilde F$ provides the appropriate brane flux for $Y$ to be coisotropic brane in $\mathcal{X}_{SU(2)}.$

In the patch $X_3\ne 0$ we may define $U={X_4\over X_3}$ and $V=WX_3^2$
so that $X_1,X_2,U,V$ describe local $\mathbb{P}^1$ with normal
bundle $\mathcal{O}(-1)\oplus \mathcal{O}(-1).$ 
In this patch
$\tilde F$
coincides with \eqref{eq:exampleIIB} for $a=b=1$ and $\lambda_1=\lambda_2=0$
which we used before to define coisotropic brane
in resolved conifold.

Meanwhile, in the patch $X_4\ne 0$ we define
$$\tilde U=WX_4^2,\quad \tilde V={X_3\over X_4}$$
so that $X_1,X_2,\tilde U,\tilde V$ describe local $\mathbb{P}^1$ geometry
with normal bundle $\mathcal{O}(-3)\oplus \mathcal{O}(1).$
In this patch $\tilde F$
coincides with \eqref{eq:exampleIIB} for $a=3,b=-1$ and $\lambda_1=1,\lambda_2=-1.$
In this way in both patches $Y$ is coisotropic brane in the corresponding local geometry.

Let us choose torus $\bf{T}$ acting on $\mathcal{X}_{SU(2)}$ as
$${\bf T}: X_1\mapsto e^{i\varphi} X_1, \quad X_2\mapsto X_2, \quad
X_3\mapsto X_3, \quad X_4 \mapsto e^{in\varphi} X_4,\quad W\mapsto e^{-i(n+1)\varphi}W.$$
Contribution to the chiral superpotential on the surface defect from disk multi-covers in both local $\mathbb{P}^1 $ geometries is

\be \label{super}\mathcal{W}=\sum_{d=1}^{\infty}{\Bigl(N_d(n)e^{-dt_1/2}+N_d(n-1)e^{-dt_2/2}\Bigr)\over d^2}e^{dy}+\sum_{d=1}^{\infty}{\Bigl(N_d(n)e^{-dt_1/2}+N_d(n+1)e^{-dt_2/2}\Bigr)\over d^2}e^{-dy}\ee
where $t_1,t_2$ are complexified K\"ahler moduli. 

As in \cite{AgVa}, there are other contributions to $\mathcal{W}$
from holomorphic maps with reducible domain $\Sigma$ that contains $k$ copies of 
$\mathbb{P}^1$  and d-multi-covers of the disk. These are suppressed at large $Re(t)$ as $e^{-(k+d/2)t}$ with appropriate $t.$
Assuming the standard K\"ahler function for $y,$ the stability of the surface defect
can be argued using \eqref{super}.
%Introducing inhomogenous coordinates
%$$\CV_1: \quad z_1={X_1\over X_2},\quad u_1=X_2^{a+2}\tilde U,\quad v_1=X_2^{-a}\tilde V$$  
%$$\CV_2: \quad z_2={X_2\over X_1},\quad u_2=X_1^{a+2}\tilde U,\quad v_2=X_1^{-a}\tilde V$$  
%we get that on a boundary of a disk i.e. at $\vert z_2 \vert=1,\,u_2=v_2=0,$
%we find boundary conditions for fermions exactly as before (\ref{bry_0_form}) and
%(\ref{bry_1_form}). 

The story is easy to generalize. Let us consider toric manifold $\mathcal{X}_{SU(N)}$
which is typically used in geometric engineering of $\mathcal{N}=2,4d$ $SU(N)$ gauge theory with $N\ge 3$
 %There are $N$ $\mathbb{C}^*$ actions
 
\be \label{action_ii}\begin{array}{c|c|c|c|c|c|c|c|c|c}
&W_1 & W_2 & W_3  & \ldots &W_{N-1}&W_N & W_{N+1}&X_1& X_2 \cr
\hline
\mathbb{C}^*_{(1)}&1 & -2 & 1& 0&0&0&0&0&0 \cr
\hline
\mathbb{C}^*_{(2)}&0& 1& -2& 1& 0&0&0&0&0\cr
\hline
\ldots&&&&\ldots &&&&0&0\cr
\hline
\mathbb{C}^*_{(N-1)}&0& 0& 0& 0 & 1& -2&1&0&0\cr
\hline
\mathbb{C}^*_{(N)}&-1&-1 & 0& 0& 0&0&0&1&1\cr
\end{array}
\ee
$M$ in this case is real codimension $N+1$ submanifold of $\mathbb{C}^{N+3}$
$$|W_1|^2-2|W_2|^2+|W_3|^2=\zeta_1, \quad |W_2|^2-2|W_3|^2+|W_4|^2=\zeta_2,
\quad \ldots,$$
$$|W_{N-1}|^2-2|W_N|^2+|W_{N+1}|^2=\zeta_{N-1},\quad |X_1|^2+|X_2|^2-|W_1|^2-|W_2|^2=\zeta_N, \quad |X_2|^2=|X_1|^2+c$$ 
where we choose $\zeta_N>c$ so that $X_1,X_2\ne 0$ on M.
$Y$ is the quotient of $M$ by symplectic $U(1)^N$ action. 

Let $X_a=r_a \,e^{i\theta_a}\, a=1,2.$
The following closed 2-form on $M$
$$\tilde F=e^{i(\theta_{1}+\theta_2)}\Biggl(\prod_{j=3}^{N+1} W_j\,
(dW_1+iW_1d\theta_1)\wedge(dW_2+iW_2d\theta_2)+$$
$$\sum_{k=3}^{N+1}\Bigl(
\prod_{j=3; \, j\ne k}^{N+1} W_j\Bigr)\,\Bigl((k-1)W_2
(dW_1+iW_1d\theta_1)+(k-2)W_1(dW_2+iW_2d\theta_2)\Bigr)\wedge dW_k
$$
$$-W_1W_2\sum_{k=3}^{N+1}\sum_{m=3}^{N+1}(k-2)(m-1)\Bigl({\prod_{j=3; \, j\ne k;\, j\ne m}^{N+1}W_j}\Bigr) \, dW_k\wedge dW_m\Biggr)$$
is invariant under symplectic $U(1)^N$ and descends to the closed  2-form $F$ on $Y.$
Similarly to the previously considered $\mathcal{X}_{SU(2)},$
$\tilde F$ provides the appropriate brane flux for $Y$ to be coisotropic brane in $\mathcal{X}_{SU(N)}.$ Namely, there are $N$ local $\mathbb{P}^1$ geometries in $\mathcal{X}_{SU(N)}.$ In each of these geometries, $\tilde F$ coincides with \eqref{eq:exampleIIB} for
the appropriate $a,b$ and $\lambda_1,\lambda_2.$

For example, in the patch $W_j\ne 0 \,j=3,\ldots,N+1$ we define
$$U={W_1\over \prod_{j=3}^{N+1}(W_j)^{j-2}},\quad 
V=W_2 \, \prod_{j=3}^{N+1}(W_j)^{j-1}$$
so that $X_1,X_2,U,V$ describe local $\mathbb{P}^1$ geometry
with normal bundle $(-1,-1).$
 In this patch $\tilde F$ coincides with \eqref{eq:exampleIIB} for $a=b=1$ and $\lambda_1=\lambda_2=0$ which we used before to define coisotropic brane
in resolved conifold. 

Let us choose $\bf{T}$ acting on $\mathcal{X}_{SU(N)}$ as
$${\bf T}: X_1\mapsto e^{i\varphi} X_1, \quad X_2\mapsto X_2, \quad
X_3\mapsto X_3, \quad \ldots \quad X_{N}\mapsto X_N,\quad, X_{N+1}\mapsto X_{N+1},$$
$$ W_1 \mapsto e^{in\varphi} \,W_1,\quad W_2\mapsto e^{-i(n+1)\varphi}\,W_2.$$

Disk multi-covers ending on coistropic brane $Y$ in each of $N$ local $\mathbb{P}^1$ geometries contribute to chiral superpotential   on the surface defect
\be \label{super_ii}\mathcal{W}=\sum_{k=0}^{N-1}\sum_{d=1}^{\infty}\Bigl({N_d(n-k)
e^{-dt_k/2}\over d^2}e^{dy}+{N_d(n+k)e^{-dt_k/2}\over d^2}e^{-dy}\Bigr)\ee
where $t_k$ for $k=0,\ldots, N-1$ are complexified K\"ahler moduli.

\vspace{0.5cm}

{\bf Acknowledgement} I am very grateful to Emanuel Diaconescu
for many helpful discussions and 
 to Jaume Gomis for inspiring my interest in coisotropic branes in toric manifolds.

\appendix 
\section{Computation of fermion zero modes}
 Let $X$ be a local $\mathbb{P}^1$ with normal bundle 
 $\mathcal{O}(-a)\oplus \mathcal{O}(-b)$ with $a+b=2.$
Let $f:\Delta \to X$ be a degree $d\geq 1$ 
torus invariant holomorphic map with coisotropic boundary 
conditions along $Y$ which factors through the embedding $D_1\subset X$. 
This appendix consists of $\check{\rm{C}}$ech cochain computations 
of cohomology spaces $H^0(\Delta, f^*\CT_{D_1,\widetilde{R_+}})$ and
 $H^1(\Delta, f^*\CT_{D_1,R_+})$ which appear in virtual localization formula
 \eqref{eq:locEulerA}.
%of the cohomology spaces $H^{i}(\Delta, \CT_\Delta)$, $H^i(\Delta, f^*\CT_{D,R_+})$, 
%$i=1,2$, defined in section (\ref{diskmulticover}).

 Let $\CV_1,\CV_2\subset \Delta$ be 
an  open cover of $\Delta$ 
so that $\CV_1=\Delta \setminus \partial_\Delta$, and $\CV_2=\Delta \setminus \{0\}$. 
Let $t_1,t_2$ be affine coordinates on $\CV_1,\CV_2$ 
so that the map $f$ is locally given by 
\be\label{eq:localmapA} 
z_1= t_1^d\qquad z_2=t_2^d.
\ee
Note that $t_1$ is centered at the origin and $t_1t_2=1$ 
on the 
overlap $\CV_1\cap \CV_2$. 

\subsection{$\chi$ zero modes}
Let us first determine $\chi$ zero modes. For this we compute $H^0\left(\Delta, 
f^*\CT_{D_1,\widetilde{R_+}}\right).$  
%with boundary conditions (\ref{bry_0_form}).
The local sections are of the form 
\[
\bal 
s_1 & = \big(\sum_{n=0}^\infty \alpha_nt_1^n \big) \partial_{z_1} + 
\big(\sum_{n=0}^\infty \beta_nt_1^n \big) \partial_{u_1} + 
\big(\sum_{n=0}^\infty \gamma_nt_1^n \big) \partial_{v_1} \\
s_2 & = \big(\sum_{n=-\infty}^\infty \alpha'_nt_2^n \big) \partial_{z_2} + 
\big(\sum_{n=-\infty}^\infty \beta'_nt_2^n \big) \partial_{u_2} + 
\big(\sum_{n=-\infty}^\infty \gamma'_nt_2^n \big) \partial_{v_2} \\
\eal 
\]
Equation (\ref{bry_0_form}) yields the following boundary conditions for $s_2$ 
\be\label{eq:boundarycondB} 
\alpha'_n + {\overline \alpha'}_{2d-n} =0 \qquad 
\beta'_n = r^{2-2b}\,{\overline \gamma'}_{-d-n}  
\ee
Moreover
\[
\partial_{z_1} = -z_2^2\partial_{z_2}\qquad \partial_{u_1} = z_2^{-a}\partial_{u_2}
\qquad \partial_{v_1} = z_2^{-b}\partial_{v_2}
\]
on the overlap $\CV_1\cap \CV_2$. 
Therefore 
\[
\bal
\delta(s_1,s_2) = &- \big(\sum_{n=0}^\infty \alpha_nt_2^{2d-n} \big) \partial_{z_2} + 
\big(\sum_{n=0}^\infty \beta_nt_2^{-ad-n} \big) \partial_{u_2} + 
\big(\sum_{n=0}^\infty \gamma_nt_2^{-bd-n} \big) \partial_{v_2} \\
& - \big(\sum_{m=-\infty}^\infty \alpha'_m t_2^m \big) \partial_{z_2} -
\big(\sum_{m=-\infty}^\infty \beta'_m t_2^m \big) \partial_{u_2} -
\big(\sum_{m=-\infty}^\infty \gamma'_m t_2^m \big) \partial_{v_2} \\
\eal 
\]
and we find zero modes of $\chi$ as $\rm{Ker}\, \delta$
\be \label{chi_zero_app}\chi^{(zm)}=\Bigl(\sum_{m=0}^{2d}\alpha'_m t_2^m \Bigr) \partial_{z_2},\quad \overline{\alpha'_m}=-\alpha'_{2d-m}.\ee

\subsection{$\psi$ zero modes}
Let us now determine $\psi$ zero modes. For this we compute $H^1\left(\Delta,
 f^*\CT{D_1,R_+}\right)).$  
%with boundary conditions (\ref{bry_1_form}).
The local sections are of the form 
\[
\bal 
s_1 & = \big(\sum_{n=0}^\infty \alpha_nt_1^n \big) \partial_{z_1} + 
\big(\sum_{n=0}^\infty \beta_nt_1^n \big) \partial_{u_1} + 
\big(\sum_{n=0}^\infty \gamma_nt_1^n \big) \partial_{v_1} \\
s_2 & = \big(\sum_{n=-\infty}^\infty \alpha'_nt_2^n \big) \partial_{z_2} + 
\big(\sum_{n=-\infty}^\infty \beta'_nt_2^n \big) \partial_{u_2} + 
\big(\sum_{n=-\infty}^\infty \gamma'_nt_2^n \big) \partial_{v_2} \\
\eal 
\]
Equation \eqref{bry_1_form} yields the following boundary conditions 
for $s_2$ 
\be\label{eq:boundarycondB_ii} 
\alpha'_n + {\overline \alpha'}_{2d-n} =0 \qquad 
\beta'_n = -r^{2-2b}\, {\overline \gamma'}_{-d-n}  
\ee
As before we use that
\[
\partial_{z_1} = -z_2^2\partial_{z_2}\qquad \partial_{u_1} = z_2^{-a}\partial_{u_2}
\qquad \partial_{v_1} = z_2^{-b}\partial_{v_2}
\]
on the overlap $\CV_1\cap \CV_2$. 
Therefore 
\[
\bal
\rm{Im} \, \delta = &- \big(\sum_{n=0}^\infty \alpha_nt_2^{2d-n} \big) \partial_{z_2} + 
\big(\sum_{n=0}^\infty \beta_nt_2^{-ad-n} \big) \partial_{u_2} + 
\big(\sum_{n=0}^\infty \gamma_nt_2^{-bd-n} \big) \partial_{v_2} \\
& - \big(\sum_{m=-\infty}^\infty \alpha'_m t_2^m \big) \partial_{z_2} -
\big(\sum_{m=-\infty}^\infty \beta'_m t_2^m \big) \partial_{u_2} +
\big(\sum_{m=-\infty}^\infty \overline{\beta'_{-m -d}}\, t_2^m \big) \partial_{v_2} \\
\eal 
\]
Zero modes of $\psi$ are 1-chains which cannot be written as $\rm{Im}\, \delta$
\be \label{zero_psi_app}\psi^{(zm)}=\sum_{k=1}^{d-1}
\Bigl(b_k t_2^{k-ad}\partial_{u_2}+c_k t_2^{k-bd}\partial_{v_2}\Bigr)\quad b_k=-r^{2-2b}\,\overline{c_{d-k}}.\ee

%For $a=b=1$ and $d>1$ these are
%$$\Bigl(\sum_{n=1,\ldots,d-1} b_{-d+n}t_2^{-d+n}\Bigr)\partial_{u_2}+
%\Bigl(\sum_{n=1,\ldots,d-1} c_{-d+n}t_2^{-d+n}\Bigr)\partial_{v_2} \quad \text{s.t.}\quad 
%c_{-d+n}\ne - \overline{b_{-n}}.$$
%We conclude that for the resolved conifold  there are $d-1$ complex zero modes of $\psi$ which can be chosen as
%$$\psi^{(zm)}=\sum_{k=1}^{d-1} \Bigl(c_{k}t_2^{-k}\partial_{v_2}+b_{k}t_2^{-k}\partial_{u_2}\Bigr)
%.$$
%If global $U(1)$ acts as
%$$X_1 \mapsto e^{i \varphi}X_1,\quad X_2 \mapsto X_2,\quad
%U \mapsto e^{i n \varphi}U,\quad V \mapsto e^{-i(n+1)\varphi }V$$
%then
%$$t_2^{-k}\, \partial_{v_2}\mapsto exp[-i\Bigl(\phi_1+\phi_V+{k\over d}(\phi_2-\phi_1)\Bigr)]\,t_2^{-k}\,\partial_{v_2}.$$

%\bibliography{adhmref.bib}
 %\bibliographystyle{abbrv}
\end{document}